\documentclass[%
 aip,
 amsmath,amssymb,
 reprint,%
]{revtex4-1}

\usepackage{graphicx}% Include figure files
\usepackage{dcolumn}% Align table columns on decimal point
\usepackage{bm}% bold math
\usepackage{colortbl}
\usepackage[dvipsnames]{xcolor}
\usepackage{makecell}
\usepackage{multirow, booktabs}
\usepackage{subcaption}
\usepackage[utf8]{inputenc}
\usepackage[T1]{fontenc}
\usepackage{mathptmx}

\newcommand{\kB}{k_\textrm{B}}

\makeatletter

    \def\CT@@do@color{%
      \global\let\CT@do@color\relax
            \@tempdima\wd\z@
            \advance\@tempdima\@tempdimb
            \advance\@tempdima\@tempdimc
    \advance\@tempdimb\tabcolsep
    \advance\@tempdimc\tabcolsep
    \advance\@tempdima2\tabcolsep
            \kern-\@tempdimb
            \leaders\vrule
                    \hskip\@tempdima\@plus  1fill
            \kern-\@tempdimc
            \hskip-\wd\z@ \@plus -1fill }
    \makeatother

\begin{document}

\preprint{AIP/123-QED}

\title[Controlling the dopant profile for SRH suppression at low current densities in $\lambda\approx$ 1330nm GaInAsP light-emitting diodes]{Controlling the dopant profile for SRH suppression at low current densities\\in $\lambda\approx$ 1330nm GaInAsP light-emitting diodes}
% Force line breaks with \\

\author{Parthiban Santhanam}
% \email{parthi@stanford.edu.}
\author{Wei Li}
% \email{weili1@stanford.edu.}
\author{Bo Zhao}
% \email{bzhao89@stanford.edu.}
\author{Chris Rogers}
% \email{}
\author{Dodd Joseph Gray Jr.}
% \email{}
\affiliation{Ginzton Laboratory, Department of Electrical Engineering, Stanford University,\\Stanford, California 94305 USA}%
\author{Phillip Jahelka}
% \email{}
%\author{Michelle Sherrott}
%% \email{}
\author{Harry A. Atwater}
% \email{}
\affiliation{Thomas J. Watson Laboratories of Applied Physics, California Institute of Technology,\\Pasadena, California 91125 USA}
\author{Shanhui Fan}
 \email{shanhui@stanford.edu.}
\affiliation{Ginzton Laboratory, Department of Electrical Engineering, Stanford University,\\Stanford, California 94305 USA}%

% \homepage{http://www.Second.institution.edu/~Charlie.Author.}
%\affiliation{%
%Second institution and/or address%\\This line break forced% with \\
%}%

\date{\today}% It is always \today, today,

\begin{abstract}
The quantum efficiency of double hetero-junction light-emitting diodes (LEDs) can be significantly enhanced
at low current density by tailoring the spatial profile of dopants to suppress Shockley-Read-Hall (SRH)
recombination. To demonstrate this effect we model, design, grow, fabricate, and test a GaInAsP LED
($\lambda\approx$ 1330nm) with an unconventional dopant profile. Compared against that of our control
design, which is a conventional $n^+$-$n$-$p^+$ double hetero-junction LED, the dopant profile near the $n$-$p^+$
hetero-structure of the new design displaces the built-in electric field in such a way as to suppress the $J_{02}$
space charge recombination current. The design principle generalizes to other material systems and could
be applicable to efforts to observe and exploit electro-luminescent refrigeration at practical power densities.
\end{abstract}

\maketitle

%\begin{quotation}
%The ``lead paragraph'' is encapsulated with the \LaTeX\ 
%\verb+quotation+ environment and is formatted as a single paragraph before the first section heading. 
%(The \verb+quotation+ environment reverts to its usual meaning after the first sectioning command.) 
%Note that numbered references are allowed in the lead paragraph.
%%
%The lead paragraph will only be found in an article being prepared for the journal \textit{Chaos}.
%\end{quotation}

% SCRAPS OF TEXT
% in accordance with 
%
%The total production costs per Watt of solar photovoltaic and solid-state lighting technologies
%have experienced logarithmic reductions in 
%
%fallen in accordance with learning curves
%
%As the cost of semiconductor opto-electronics in the solar photovoltaic and lighting
%industries has fallen, the technology platforms which share capital base with them
%have become serious contenders for
%scalable new energy technologies\cite{ITRPV10}\cite{Haitz2000}\cite{Steele2007}\cite{Haitz2011}\cite{Gerke2015}.
%As the price per area of semiconductor opto-electronics in the solar photovoltaic, lighting,
%and display industries has fallen, the technology platforms which share capital base with them
%have become serious contenders for scalable new energy technologies. 

As the solar photovoltaic and solid-state lighting industries have continued to deliver
reductions in both the cost per semiconductor area and cost per Watt of their energy technology
products \cite{ITRPV10,Haitz2000,Steele2007,Haitz2011,Gerke2015},
the technology platforms which share capital base with them have become serious contenders for
scalable new energy technologies. Two of the most durable sources of demand for energy conversion
which have yet to be ``semiconductorized'' are air conditioning and waste heat
recovery. Both of these demands have potentially viable solutions utilizing III-V
opto-electronics\cite{Xiao2019,Heikkila2009,Piprek2016,Chen2017,Harder2003,Zhao2018}.
However it is also true that high-efficiency realizations of opto-electronic devices that
can in theory meet these needs often carry the same stringent requirements on quantum efficiency
as do Light-Emitting Diodes (LEDs) utilizing electro-luminescent cooling to act as refrigerators.

As a result, despite multiple recent demonstrations of photonic heat flux control capable
of solid state cooling\cite{Hehlen2018,Zhu2019,Radevici2019}, the pace of
progress on these potential new energy technologies has been limited by the
technical challenges of fabricating devices with high enough quantum efficiencies for cooling
at practical power densities. The general problem is two-fold: first electron-hole pairs injected
into the luminescing material must be efficiently converted into photons within that material,
and second the photons thus generated must be able to escape the high refractive index typical
of the luminescent materials which allow the first point to be satisfied.

%.bib file previously had this element included
%@ARTICLE{Wang2018,
%	author = "Matthew S. Wang and David Hwang and Abdullah I. Alhassan and Changmin Lee and Ryan Ley and Shuji Nakamura and Steven P. DenBaars",
%	title = "High wall-plug efficiency blue III-nitride LEDs designed for low current density operation",
%	journal = "Optics Express",
%	volume = 26,
%	number = 16,
%	month = aug,
%	year = 2018,
%	doi = "10.1364/OE.26.021324"
%}
%

Serious progress in overcoming both of these challenges has been made experimentally in the past few years
\cite{Radevici2019,Kuritzky2017,Li2019}. The light extraction efficiency challenge has
been addressed in at least two distinct ways. First, a recent experiment
redesigned LED structures \cite{Kuritzky2017} in the well-developed blue Nitride LED material system
for low current density where thinner current-spreading layers which significantly reduce the probability
that recycled photons will be lost to irreversible absorption in contact metal can be sufficient.
Second, but no less relevant, GaAs photo-diodes
that served as photon-absorbing detectors were grown in series with GaAs emitters in the same epitaxial
layer stack thereby enabling direct light extraction from the LED to the photodetector, which is more efficient
as compared with light extraction from an LED to free space\cite{Radevici2019}.

The other half of the technical challenge, to increase the fraction of injected electron-hole pairs
which recombine radiatively (i.e. the Internal Quantum Efficiency or IQE to the light-emitter
community; alternatively the Internal Radiative Efficiency or IRE to the photo-voltaic community),
has recently been approached by attempts to suppress parasitic SRH recombination that prevents
cooling from being observed at lower forward bias voltages, where the problem of light extraction
is more forgiving. Operating an LED at a slightly lower forward bias voltage than typical
results in a larger ratio between the so-called photon voltage $V_p \equiv \frac{\hbar\omega}{q}$,
where $\hbar\omega$ is the emitted photon's energy and $q$ is the charge of an electron, and the applied
voltage $V$. When the ratio $V_p/V$ is greater than unity but by a small margin, as tends to be
the case in blue Nitride LEDs (even at current densities considered low relative to most practical
devices), the margin for error is correspondingly small. A ratio of $V_p/V=1.03$, for instance,
implies that no more than 3\% of the internally generated photons may be lost to parasitic absorbing
structures (e.g. contacts) without entirely eliminating the possibility of net cooling. A ratio
of $V_p/V=1.30$, however, increases this margin for error ten-fold, making experimental realization
drastically more tractable. While this approach does sacrifice power density, it makes observation of a
large steady-state temperature reduction of the emitting device less demanding and could enable
photonic solid state coolers that support large temperature differences with very small cold-side
heat loads in the near future.

\begin{figure*}
\begin{subfigure}{0.31\textwidth}
\includegraphics[width=\linewidth]{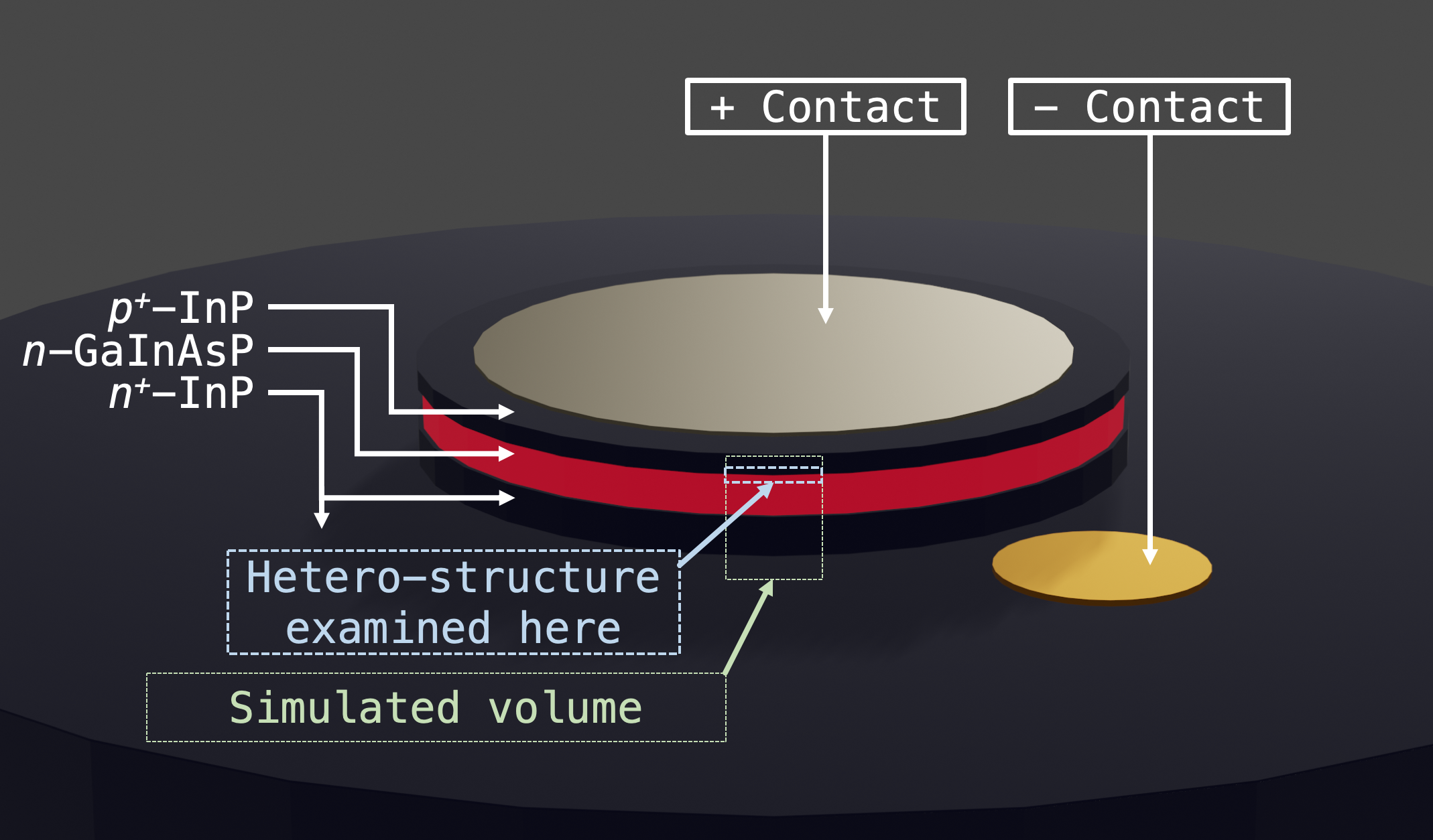}
\caption{Graphical depiction.} \label{fig:depiction}
\end{subfigure}
\hspace*{\fill} % separation between the subfigures
\begin{subfigure}{0.31\textwidth}
\includegraphics[width=\linewidth]{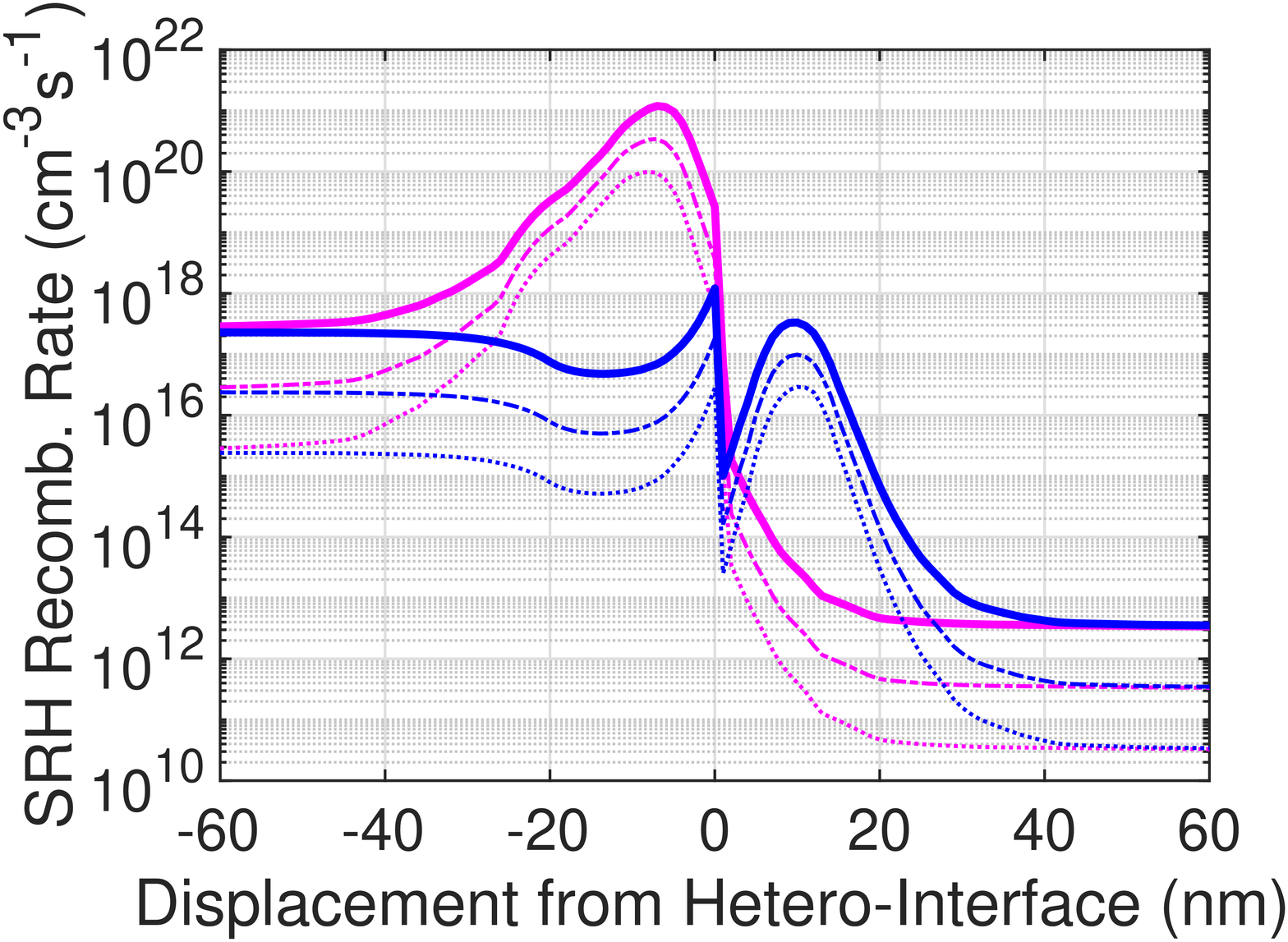}
\caption{SRH profiles.} \label{fig:DesignSRHprofile}
\end{subfigure}
\hspace*{\fill} % separation between the subfigures
\begin{subfigure}{0.31\textwidth}
\includegraphics[width=\linewidth]{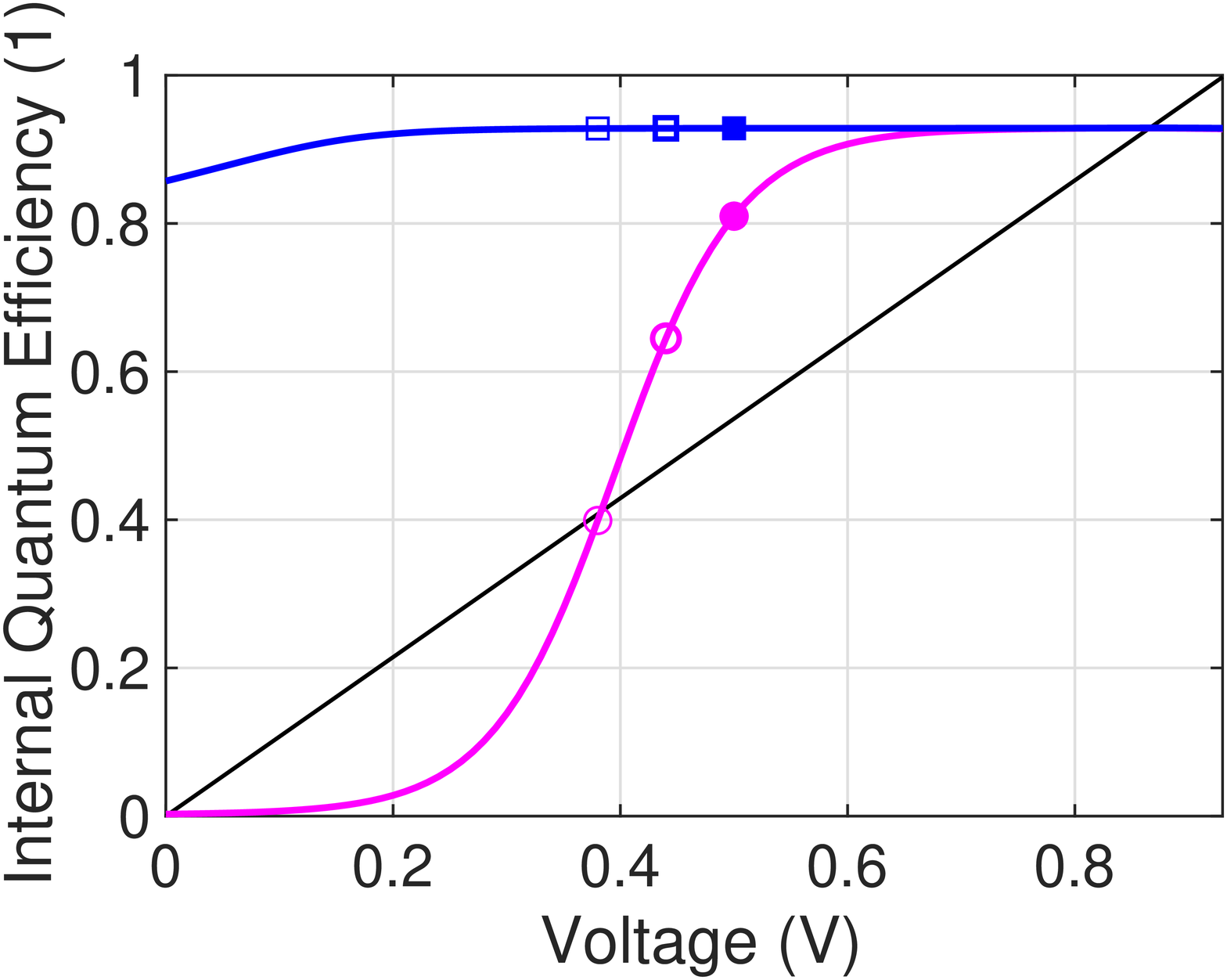}
\caption{Low-voltage IQE enhancement.} \label{fig:DesignIQEvsV}
\end{subfigure}
\captionsetup{justification=raggedright}
\caption{In (a), a depiction of the fabricated devices which highlights the $n$-$p^+$
hetero-structure examined here. In (b), the spatial distribution of
bulk trap-assisted SRH recombination is shown to have its peak displaced between the conventional
control design (magenta) and the design for SRH suppression presented here (blue). The solid
lines represent the recombination rate profile at 500 mV, while the dash-dotted
and dotted lines represent the same devices at 60 and 120 mV lower bias respectively. Displacements from
the hetero-interface obey the convention that negative values are in the smaller bandgap $n$-GaInAsP while
positive values are in the larger bandgap $p$-InP. In (c), the consequences
of the design change for IQE are plotted as a function of voltage. Again the two designs
are distinguished by color. Note that the minimum quantum efficiency for which net electro-luminescent
refrigeration is theoretically possible for ideal photon extraction is denoted by the straight diagonal line.} \end{figure*}

Recent work has produced a near-infrared LED with suppressed SRH recombination
at the values of $V_p/V$ most relevant to electro-luminescent cooling\cite{Li2019}.
In Ref.~\onlinecite{Li2019}, the authors designed, fabricated, and tested an LED which made use of the spectacularly
low interfacial surface recombination velocity at the hetero-interface between the luminescent
material (1.3eV-GaInAs) and the wider bandgap carrier confining material (GaInP) to extend the
high IQE of near-infrared LEDs to much lower current densities where
non-radiative Shockley-Read-Hall recombination is usually dominant. The structure in Ref.~\onlinecite{Li2019}
utilizes a very thin GaInAs quantum well to confine the injected carriers to a narrow interval
in the growth direction where they experience non-radiative Shockley-Read-Hall (SRH)
recombination a very small fraction of the time. Thus they observe very high IQE from their LEDs even at
forward biases $\approx$200 mV below the emitted photon energy. This strategy for increased IQE,
however, comes at the price of current density as only electron-hole pairs in this same narrow interval can
contribute to radiative recombination. As a result, the current density achieved in Ref.~\onlinecite{Li2019} is
more than two orders of magnitude below the detailed-balance limit for an opaque body of the same bandgap.
In pursuit of applications with more moderate current densities, such as most cooling experiments to date,
some of this may be addressed by the use of multiple quantum wells or much thicker emitting layers but
whether such structures can retain high IQE performance remains to be seen.

%\begin{figure*}
%\includegraphics[width=0.28\linewidth]{./FIG_SCRIPT_OUTPUTS/Blender/Diagram_of_Mesa_Fig1_v3.png} $\quad\quad$
%\includegraphics[width=0.28\linewidth]{./FIG_SCRIPT_OUTPUTS/Lumerical/plot_AvsB_SRH_profile.eps} $\quad\quad$
%\includegraphics[width=0.28\linewidth]{./FIG_SCRIPT_OUTPUTS/Lumerical/plot_AvsB_IQE_of_V.eps}
%\caption{\label{fig:SRHprofile} At left, a depiction of the fabricated devices which highlights the $n$-$p^+$
%hetero-structure examined in this Letter. At center, the spatial distribution of trap-assisted recombination is shown to have
%its $J_{02}$ peak displaced between the conventional control design (magenta) and the design for SRH suppression presented here
%(blue). In this plot, the solid lines represent the recombination rate profile at 500 mV, while the dash-dotted
%and dotted lines represent the same devices at 60 and 120 mV lower bias respectively. Displacements from
%the hetero-interface obey the convention that negative values are in the smaller bandgap $n$-GaInAsP while
%positive values are in the larger bandgap $p^+$-InP. At right, the consequences
%of the design change for internal quantum efficiency are plotted as a function of voltage. Again the two designs
%are distinguished by color. Note that the minimum quantum efficiency for which net electro-luminescent
%refrigeration is theoretically possible even for lossless photon extraction is denoted by the straight diagonal line.}
%%connecting the origin (0,0) to ($\frac{\langle\hbar\omega\rangle}{q}$,1)}.
%\end{figure*}

In this work we aimed to improve IQE at deep-sub-bandgap bias voltages through a distinct
SRH suppression strategy, alluded to but not fully examined in Ref.~\onlinecite{Gray2013},
which utilizes a micron-thick emitting layer so that even in a device realization
with near-complete SRH suppression, the diode remains near the detailed balance limit
for current density. We report work on GaInAsP LEDs, grown lattice-matched to their InP substrates and
emitting photons at a free-space wavelength of $\lambda \approx$ 1330nm, whose epitaxial layer stack design
aims to extend downward the range of forward bias voltages where high IQE is observed by the suppression
of bulk non-radiative SRH recombination. We subsequently fabricate devices resembling Fig.~\ref{fig:depiction} and characterize them.
Our approach targets the bulk SRH recombination in the electrically
non-neutral space charge region around the $p$-$n$ junction (i.e. the often-parasitic $J_{02}$ current \cite{Grove1967}
contribution that causes the $J$-$V$ curve to exhibit a diode ideality factor of 2) of a conventional
near-infrared LED with a micron-scale emitting layer. It does so in a way that is readily generalized
to other promising material systems including GaAs.

By inserting one higher-doped layer into the $n$-GaInAsP and one lower-doped layer into the $p$-InP
regions directly surrounding the $p$-$n$ junction, the spatial distribution of bulk SRH can
theoretically be displaced enough to place its peak in the wider bandgap material. As we show presently,
the aggregate rate at which carriers undergo SRH recombination integrated over the volume of the device
is thereby reduced by orders of magnitude.

%\section{Designing the Dopant Profile for SRH Suppression}
\vspace*{0.1cm}
\noindent \emph{Design.}
The epitaxial layer stack designs on which the experimental results of this Letter are based are shown
in Table~\ref{table-layerstacks} with the layers modified to achieve SRH suppression highlighted.
Design $\mathcal{A}$ includes the proposed suppression layers while Design $\mathcal{B}$ serves as a control
against which to evaluate Design $\mathcal{A}$.

{\renewcommand{\arraystretch}{1.2}
\begin{table}[]
%\begin{minipage}{\textwidth}
%\centering
\begin{tabular}{ c|c|c|c }
%\hline
%\color{red!50!green!20!blue}
\rowcolor{Thistle!0}
\begin{tabular}{@{}c@{}} Thickness \\ (nm) \end{tabular} \
	& \hspace*{5mm}Material\hspace*{5mm} \
	& \begin{tabular}{@{}c@{}} Doping in $\mathcal{A}$ \\ $\mid$$N_D-N_A$$\mid$ (cm$^{-3}$) \end{tabular}	\
	& \begin{tabular}{@{}c@{}} Doping in $\mathcal{B}$ \\ $\mid$$N_D-N_A$$\mid$ (cm$^{-3}$) \end{tabular}	\\ \midrule[1.2pt]
480 \
	& $p$-InP	\
	& 1$\times 10^{18}$	\
	& 1$\times 10^{18}$	\\ \hline
20
	& $p$-InP \
	& \cellcolor{blue!3}1$\times 10^{17}$	\
	& \cellcolor{blue!10}1$\times 10^{18}$	\\ \hline
20
	& $n$-GaInAsP
	& \cellcolor{blue!30}2.26$\times 10^{18}$	\
	& \cellcolor{blue!10}4$\times 10^{17}$	\\ \hline
980
	& $n$-GaInAsP
	& 4$\times 10^{17}$	\
	& 4$\times 10^{17}$	\\ \hline
50
	& $n$-InP
	& 4$\times 10^{17}$	\
	& 4$\times 10^{17}$	\\ \hline
1250
	& $n$-InP
	& 2$\times 10^{18}$	\
	& 2$\times 10^{18}$	\\ \hline
200
	& $n$-InP
	& 1$\times 10^{18}$	\
	& 1$\times 10^{18}$	\\ \midrule[1.2pt]
$\approx$370$\mu$m \rule{0pt}{2.5ex}
	& $i$-InP
	& Semi-Insulating \
	& Semi-Insulating \\ \midrule[1.2pt]
\end{tabular}
\captionsetup{justification=raggedright}
\caption{\label{table-layerstacks}Epitaxial layer stacks as designed prior to growth.
Design $\mathcal{A}$ is a modified double hetero-junction epitaxial layer stack which utilizes
an intentional inhomogeneous doping profile to displace the depletion region out of the
GaInAsP and into the adjacent InP layer. Design $\mathcal{B}$ is a conventional
double hetero-junction epitaxial layer stack.}
%\end{minipage}
\end{table}}

To explain the design process and clarify the intended generalizations of the design principle into
other material systems, we focus on the electron transport in the region surrounding the $p$-$n$
hetero-junction buried 500 nm below the surface in each design (i.e. where the two designs differ).
We seek to model the transport in this region in the case of a forward bias voltage $V$ that is
below the band gap energy's voltage scale $E_\textrm{gap}/q$ by several multiples of the
thermal voltage $V_\textrm{th}$ ($V_\textrm{th} \equiv \kB T/q$, where $\kB$ is Boltzmann's constant,
$T$ is the absolute temperature of the semiconductor lattice, and $q$ is the magnitude of the electron's charge).
This voltage regime has been previously referred to\cite{Santhanam2015} as the ``deep sub-bandgap'' bias regime
and is tens to hundreds of milli-volts below the bias voltages where LEDs are typically designed to operate.
At these deep sub-bandgap operating points, a very significant fraction of the current which
flows through the device corresponds to leakage through trap-assisted SRH recombination.

Furthermore, the standard formulation of SRH recombination indicates that the spatial distribution
of SRH recombination is highly localized around a particular plane normal to the growth direction.
In this Letter, we refer to this plane as the Active Trap plane and denote its position as $z = z_\textrm{AT}$.
Here we have used the standard formulation\cite{Hall1952,Shockley1952,Grove1967} for the %Note: In Grove1967: Section 5.2, Equation 5.41
local rate of trap-assisted Shockley-Read-Hall (SRH) recombination as expressed in the following equation:
\begin{equation} \label{eq:SRHrate}
W_\textrm{SRH} = \frac{\Big( \textrm{e}^\frac{qV}{\kB T} - 1 \Big) \cdot n_i^2}{(n+n_i)\tau_p + (p+n_i)\tau_n} \quad\quad\textrm{.}
\end{equation}
%\begin{equation} \label{eq:SRHrate}
%W_\textrm{SRH} = \frac{\left( \textrm{e}^\frac{qV}{\kB T} - 1 \right) \cdot n_i^2}{(n+n_1)\tau_p + (p+p_1)\tau_n}
%\end{equation}
%\begin{equation} \label{eq:SRHrate}
%W_\textrm{SRH} = \frac{(np-n_i^2)}{(n+n_1)\tau_p + (p+p_1)\tau_n}
%\end{equation}
%where $n_1$ and $p_1$ are defined as

The above equation assumes the trap spectrum is a delta function and nonzero only
at the intrinsic Fermi level of each semiconductor. If we further assume that
the SRH lifetimes of the electrons and holes are equal (i.e. $\tau_n = \tau_p = \tau$),
we may further simplify the denominator of Equation \ref{eq:SRHrate} to find that
\begin{equation} \label{eq:SRHpeak_vs_np}
W_\textrm{SRH}(z) \propto \frac{1}{n(z) + p(z) + 2n_i(z)} \quad\quad\textrm{.}
\end{equation}
Since $n$ and $p$ are varying in $z$ and away from the hetero-junction itself that variation
is monotonic, if we temporarily assume the removed parts of Equation \ref{eq:SRHrate} are not varying
spatially, we should expect a local maximum to $W_\textrm{SRH}(z)$ around the plane
where $n+p$ is minimized. Since in quasi-equilibrium, the product of the electron and hole
concentrations $np$ is fixed for a fixed Fermi level separation, this minimum is located where $n$=$p$.
%Note that the aforementioned observation of a constant $np$ product is sometimes referred to
%as the law of mass action.
This defines $z_\textrm{AT}$: $W_\textrm{SRH}(z)$ has a local maximum at $z_\textrm{AT}$.

Fig.~\ref{fig:DesignSRHprofile} depicts the results of numerical simulations conducted with
the commercial electron transport CHARGE solver within Lumerical Device\cite{Lumerical}. We find that
the net recombination rate per unit volume near $z_\textrm{AT}$
for Designs $\mathcal{A}$ and $\mathcal{B}$ side-by-side.
Design $\mathcal{A}$ (i.e. blue curves) has its $J_{02}$ peak 10 nanometers away
from the hetero-interface into the wide-bandgap $p^+$-InP. Design $\mathcal{B}$ (i.e. magenta
curves) has its $J_{02}$ peak $\approx$7 nanometers away from the hetero-interface into the
narrower-bandgap $n$-GaInAsP.

Due to the strength of the built-in electric field in this region and our assumptions about the trap spectrum,
this local maximum is fairly sharp. For operating points in the deep sub-bandgap bias regime, a field strength
of order 10$^7$ V/m is typical. In this case we may use the room temperature rule of thumb of 60 meV/decade
for the exponential decay of the concentration of electrons and holes as a function of voltage in the Boltzmann limit
at $\approx$300 K to quantify this sharpness. For the assumed field strength, a plane corresponding to $z$ just
6 nanometers away from $z_\textrm{AT}$ would have either $n$ or $p$ a full ten times the value of
$n(z_\textrm{AT}) = p(z_\textrm{AT})$. Since both $n$ and $p$ enter the denominator additively
in Equation \eqref{eq:SRHpeak_vs_np}, this indicates that $\approx$99\% of the integrated recombination
takes place within a 25 nm window around $z_\textrm{AT}$. For the admittedly unphysical case of a perfect delta
function trap spectrum, this corresponds to more than half of the recombination being localized within
a window of width 50 \text{\normalfont\AA}.

Temporarily neglecting the additional complication of the hetero-junction, the physical interpretation
of this peak is reasonably simple. Outside of this window, we find the traps are less active. If we
consider planes displaced in the growth direction toward the $n$ terminal of the device, we find
that the hole concentration $p$ is too low there and so the SRH process is bottlenecked by slow
hole capture rates. Similarly, in planes displaced in the growth direction toward the $p$ terminal of the
device, we find that the electron concentration $n$ is too low and the SRH process is bottlenecked by slow
electron capture rates.

For our control structure in Design $\mathcal{B}$, this peak falls within the narrow bandgap GaInAsP active region.
Our goal in Design $\mathcal{A}$ is to displace this spatial peak in SRH rate into the wider gap InP material
so that the $n_i^2$ term in the numerator of Equation \eqref{eq:SRHrate} can suppress it. To do this
we will modify the dopant impurity profile from that in Design $\mathcal{B}$ to that in Design $\mathcal{A}$
as shown in Table~\ref{table-layerstacks}.

%\section{Fabrication\label{sec:Fab}}
\vspace*{0.1cm}
\noindent \emph{Fabrication.}
Custom epitaxial growths were purchased commercially from OEpic, Inc. in Sunnyvale, CA.
Growth was done by MOCVD on Fe-doped semi-insulating InP substrates.
The growths reported here were all done during a pair of back-to-back
runs of the same reactor with the intention of eventually comparing device
characteristics in a controlled manner.
%
%\begin{figure}
%\includegraphics[width=0.8\linewidth]{./FIG_SCRIPT_OUTPUTS/PL/pl_spectra_cryo_and_room_temperature.png}
%\captionsetup{justification=raggedright}
%\caption{\label{fig:PL} Photo-luminescence spectra of the bare epitaxy. Measurements from room temperature
%($\approx$300 K) appear in red and measurements from low temperature ($>$4 K) appear in blue.}
%\end{figure}

\begin{figure*}
\begin{subfigure}{0.4\textwidth}
\includegraphics[width=\linewidth]{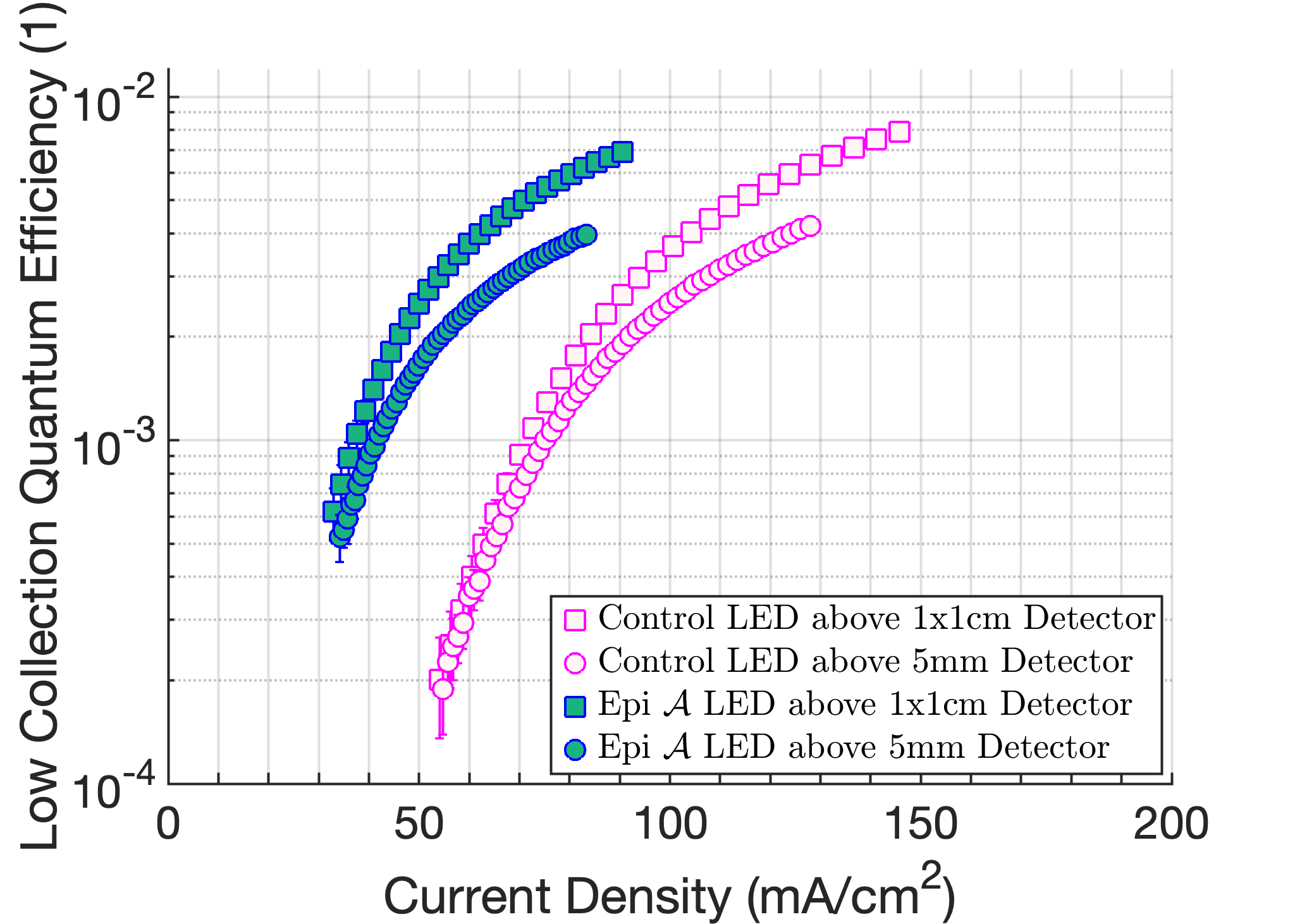}
\caption{Quantum Efficiency at Photo-Detector.} \label{fig:LIV}
\end{subfigure}
\hspace*{1cm} % separation between the subfigures
\begin{subfigure}{0.4\textwidth} \label{fig:characterization:IVandPL}
\includegraphics[width=\linewidth]{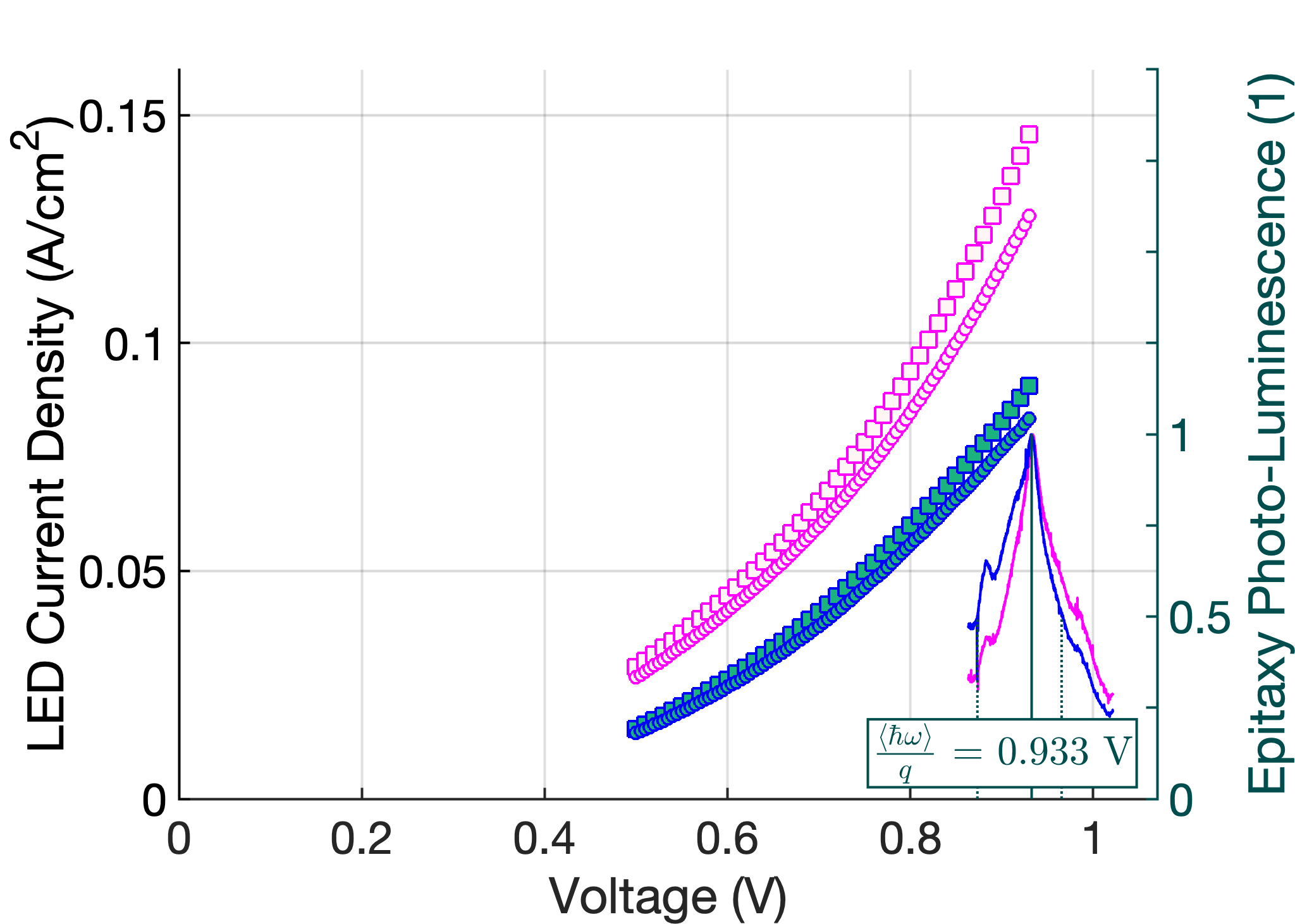}
\caption{Device $J$-$V$ and Epitaxy PL.} \label{fig:IVPL}
\end{subfigure}
\captionsetup{justification=raggedright}
\caption{\label{fig:characterization}In (a), two 8 mm$^2$ LEDs, both emitting $\lambda\approx$ 1330nm light, were successively placed about 1 mm above
each of two Germanium photo-diode detectors. The LEDs' measured external quantum efficiency was reduced moderately by this choice of spacing, but
effects from inconsistent optical coupling between emitter and detector were thereby removed. The ratio of the photo-current in the detector to
the current supplied to the LED is described as the ``Low Collection Quantum Efficiency'' and is interpreted as a measure of the true overall
quantum efficiency multiplied by some unknown low collection efficiency fraction.
The empty magenta markers denote Design $\mathcal{B}$.
The solid blue-green markers denote Design $\mathcal{A}$, although SIMS measurements
indicate the achieved dopant incorporation at the 20nm highly doped GaInAsP layer deviated significantly from the
target concentration. In (b), the $J$-$V$ data is plotted alongside the relative intensity PL spectra from the epitaxial structures
corresponding to their color.}
\end{figure*}

The first set of experiments done to verify that the films had been grown as intended
were Photo-Luminescence (PL) measurements on the bare epitaxy. The PL spectra at
room temperatures are shown later on the right vertical axis of subfigure (b) in Fig.~\ref{fig:characterization}.
%The room temperature measurement was done in collaboration with Dr. Tomas Sarmiento
%in the lab of Prof. Jelena Vuckovic. 
Cryogenic PL measurements were also done on samples which had undergone a mesa etch
but no further processing.
% and were conducted in
%collaboration with Dr. Chris Rogers and Dr. Dodd Gray in the lab of Prof. Hideo Mabuchi.
The cryogenic measurements were taken in a chamber cooled by liquid helium but
due to imperfect heat sinking in the sample mounting solution, the exact temperature
of the sample was highly uncertain. The primary conclusion of the PL experiments was
that the devices fabricated from this epitaxy should emit light with a free space
wavelength of around $\lambda\approx$1330nm.

\begin{figure}
\includegraphics[width=0.7\linewidth]{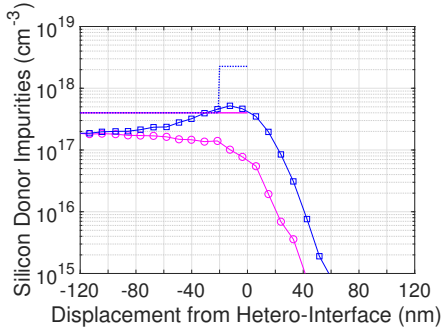}
\captionsetup{justification=raggedright}
\caption{\label{fig:SIMS} Secondary Ion Mass Spectrometry (SIMS) was used to investigate the bare, unprocessed epitaxy grown by MOCVD.
The discrete markers denote experimental data while the marker-free lines represent the dopant profile as specified in the design.
As in other figures, magenta refers to Design $\mathcal{B}$ while blue refers to Design $\mathcal{A}$.}
%Note that the increased
%concentration of silicon donors found on the narrow-gap side of the hetero-junction is significantly below the concentration
%specified in the design.}
\end{figure}

The second set of experiments was aimed at verifying the distinction between the films
grown to meet the two different design specifications. Secondary Ion Mass Spectrometry (SIMS)
analysis was done commercially by Evans Analytical Group. % in March of 2018 and was filed internally with EAG as Job Number C0JYZ934
The primary conclusion drawn from the SIMS data shown in Fig.~\ref{fig:SIMS}
was that the intended silicon donor concentration in the emitting-bandgap SRH
suppression layer was significantly higher than in the corresponding layer of the control sample,
but was far enough short of the design to create a much smaller change in the IQE of resulting LEDs
than is suggested by Fig.~\ref{fig:DesignIQEvsV}. As shown in Fig.~\ref{fig:SIMS}, the additional
silicon donors detected by SIMS were localized to a volume $\approx$60 nm thick in the growth direction
rather than the 20 nm in the design. Moreover the $z$-integrated density of silicon donors
in the sample grown targeting Design $\mathcal{A}$ exceeded that of the sample grown targeting
Design $\mathcal{B}$ by $\approx$1.8$\times$10$^{12}$ cm$^{-2}$ as compared against an excess
donor density of $\approx$3.7$\times$10$^{12}$ cm$^{-2}$ anticipated based on the designs.
Unfortunately neither the actual concentrations of free carriers nor the ionization fraction
of the silicon dopants has yet been measured directly in any part of the epitaxial films.

Next, diodes were fabricated from these films. We utilized the photo-lithography capabilities
of the tools managed by the Stanford Nanofabrication Facility and the Stanford Nano Shared Facilities,
with particular reliance on the Heidelberg Maskless Aligner MLA150, Shipley S1813 photo-resist,
and Metal-Free developer MF319. All masks were designed manually with Python scripts used to generate CIF files.
Etching was done by a combination of 3:1 H$_3$PO$_4$:HCl, which selectively etched the InP,
and 1:1:10 H$_2$O:H$_2$O$_2$:H$_2$SO$_4$, which selectively etched the GaInAsP quaternary.
The contacts were formed by lift-off processes utilizing the
same resists as for the etch masks and a pair of deposition tools. The $n$-type contacts were
simple Ti/Pt/Au films deposited using electron beam evaporation.
The $p$-type contacts were annealed Pd/Zn/Pd. Prior to the $p$-metal contacts
reported here, a generation of devices was made using simple Ti/Pt/Au for both contact
types and, due to the absence of a epitaxial contact layer, the resulting Schottky barriers
prohibited meaningful $I$-$V$ and $L$-$I$ characterization.

%\begin{figure*}
%\includegraphics[width=0.46\linewidth]{./fabrication/SEM_and_other_imaging/NovaSEM/2018.09.27/1330_Gen2_100x.jpg} $\quad\quad$
%\includegraphics[width=0.46\linewidth]{./fabrication/SEM_and_other_imaging/Keyence_Confocal/1330_Gen2_and_Gen4/Gen2_mesa.jpg}
%\captionsetup{justification=raggedright}
%\caption{\label{fig:mesa}At left an SEM. At right a reconstructed topographical map with exaggerated vertical displacements for ease of
%visibility.}
%\end{figure*}

Two types of diode structures were fabricated and tested.  Small circular mesa type diodes with
an active emitting area of 0.1 mm$^2$ each were designed for high current density. Larger square
mesa diodes were designed for lower current density measurements, each having an active
emitting area of 8 mm$^2$.

%\section{Device Characterization\label{sec:LIV}}
\vspace*{0.1cm}
\noindent \emph{Characterization.} Finally the characteristic Current-Voltage (i.e. $I$-$V$) and
Light-Current (i.e. $L$-$I$) behaviors of the devices were measured. Imperfections in various
stages of the fabrication process led to some variation in device behavior both within a given die
and from die-to-die. Particularly common were parasitic shunt pathways presumed to arise from flashes
of contact metal unintentionally attached to the surface of the devices amid
the lift-off contact fabrication process.

%\begin{figure}
%\includegraphics[width=\linewidth]{./FIG_SCRIPT_OUTPUTS/LIVT/Figure_LIV_4_sets.png}
%\captionsetup{justification=raggedright}
%\caption{\label{fig:wwasher} Two 8 mm$^2$ LEDs, both emitting $\lambda\approx$ 1330nm light, were successively placed about 1 mm above
%each of two Germanium photo-diode detectors. The LEDs' measured external quantum efficiency was reduced moderately by this choice of spacing, but
%effects from inconsistent optical coupling between emitter and detector were thereby removed. The empty magenta markers denote
%data from an LED fabricated from epitaxy grown to meet the specification from Design $\mathcal{A}$. The solid blue-green markers denote
%data from an LED fabricated from epitaxy grown with the target doping profile from Design $\mathcal{B}$, although SIMS measurements
%indicate the achieved dopant incorporation at the highly doped ``emitting bandgap depletion layer'' deviated significantly from the
%target concentration.}
%\end{figure}

Experiments were performed primarily near room temperature, measured at about 295 K. The fabricated diodes
were placed in a variety of geometries near a 1 cm$^2$ ThorLabs FDG10X10 large-area free-space Germanium photo-diode
as well as a smaller ThorLabs FDG05 Germanium photo-diode of 5 mm diameter. These photo-current from these
detectors was then amplified by a ThorLabs PDA200C TIA to measure the electro-luminescence signal.
Configurations in which the substrate of the LED suffered from limited repeatability which we ascribe to
dust and other debris, a hypothesis supported by significant increases in the signal seen when high-index
immersion oil was used at the interface. For the sake of maximal repeatability, the measurements we
present here were taken when the emitting diode chips were placed atop a nylon washer at a height about
1 mm above the photo-sensitive surface of each of the two detectors described above. The devices were
probed in a manual probe station with shielded electrical connections to a Keithley 2635B Sourcemeter.

Figure \ref{fig:characterization} presents data taken on the large 8 mm$^2$ diodes. It indicates that the epitaxial
layer stack design change from Design $\mathcal{B}$ to Design $\mathcal{A}$ resulted in less current leaking
through the device in the deep sub-bandgap bias regime as intended. The measurements are compatible with
a moderate suppression of non-radiative recombination at low current densities, leading to improved
quantum efficiency at these operating points.

%Note: Consider including sample numbers here.

%\section{Conclusions\label{sec:conc}}
\vspace*{0.1cm}
\noindent \emph{Conclusions.}
The series of experiments reported here provides significant evidence that a rational design
of the dopant profile of a near-infrared LED represents a viable technical de-risking
trajectory for the development of novel energy conversion devices based on the net refrigeration
of an LED at bias voltages $\frac{1}{2}(E_\textrm{gap}/q) < V < (E_\textrm{gap}/q - 3 \kB T)$.

%\begin{figure}
%\includegraphics[width=\linewidth]{./FIG_SCRIPT_OUTPUTS/Blender/Diagram_of_Mesa_Fig1_v3.png}
%\caption{\label{fig:diagram} Diagram of a typical fabricated  $\lambda\approx$1300nm GaInAsP LED highlighting the
%hetero-structure at which the design presented here differs from the control design.}
%\end{figure}
%

\begin{acknowledgments}
Dr. Ping-Show Wong of OEpic performed the MOCVD epitaxial growth.
Room temperature PL spectra were acquired in collaboration with Dr. Tom\'{a}s Sarmiento
in the lab of Prof. Jelena Vuckovic.
%Cryogenic PL spectra were acquired in
%collaboration with Dr. Chris Rogers and Dr. Dodd Gray in the lab of Prof. Hideo Mabuchi.
Tom Carver deposited the $n$-metal Ti/Pt/Au.
%Phil Jahelka deposited the $p$-metal Pd/Zn/Pd upon the suggestion of
Dr. Michelle Sherrott helped diagnose an early $p$-metal issue.
Prof. Eric Pop provided advice and aid in characterization facilities access.
The selective wet etch recipes were based on suggestions from Dr. Vijay Jayaraman of Praevium Research, Inc.
%There are many people to be acknowledged but Prof. Shanhui Fan was the person whose patience was probably most rigorously tested.
This work was supported by
the US Department of Energy ``Light-Material Interactions in Energy Conversion''
Energy Frontier Research Center under grant no. DE-SC0001293 and
the US Department of Energy ``Photonics at Thermodynamic Limits''
Energy Frontier Research Center under grant no. DE-SC0019140.

\end{acknowledgments}
\nocite{*}
\bibliography{bib_santhanam_apl_2019}% Produces the bibliography via BibTeX.

\end{document}